\documentclass[3p, 12pt, onecolumn]{elsarticle}
\usepackage{graphicx,amscd,amsmath,amssymb,verbatim}
\usepackage{amsfonts,epsfig}
\usepackage{mathptmx}
\usepackage{setspace}
\usepackage{epsfig}
\usepackage{url}
\usepackage{multirow}
\usepackage{subfigure}
\usepackage{color}

\begin{document}

\journal{Elsevier}

\begin{frontmatter}

\title{A simple study of the correlation effects in the superposition of
waves of electric fields: the emergence of extreme events}

\author{Roberto da Silva, Sandra D. Prado}

\address{Instituto de F{\'i}sica, Universidade Federal do Rio Grande do Sul, 
Av. Bento Gon{\c{c}}alves, 9500 - CEP 91501-970, Porto Alegre, Rio Grande do Sul, Brazil}

\begin{abstract}

In this paper, we study the effects of correlated random phases in the
intensity of a superposition of $N$ wave-fields.  Our results suggest that 
regardless of whether the phase distribution is continuous or discrete if the 
phases are random correlated variables, we must observe a heavier tail distribution 
and the emergence of extreme events as the correlation between phases increases. 
We believe that such a simple method can be easily applied in other situations to 
show the existence of extreme statistical events in the context of nonlinear complex 
systems.

\end{abstract}

\end{frontmatter}

\setlength{\baselineskip}{0.7cm}

\section{Introduction}

\label{Section:Introduction}

Rogue or freak waves appear in the deep sea. They have very high amplitudes
when compared with the surrounding waves \cite{Onorato}, but these
\textquotedblleft monsters of the sea\textquotedblright\ do not only appear
in the ocean. Rogue waves have appeared in a variety of fields from
photonics \cite{Dudley2014} to Economy \cite{YZhenYa2010}, where a nonlinear
wave model is a good alternative to the Black-Scholes model. Some authors
have explored the emergence of such extremes events comparing situations
occurring in hydrodynamics and optics \cite{Koussaifi}

The history begins with Rayleigh studies on the superposition of amplitudes
of harmonic oscillations in the 1880s. He concluded that the distribution of
such amplitudes was given by (probability density function) PDF 
\begin{equation*}
f(x)=\frac{x}{\sigma ^{2}}\exp \left( -\frac{x^{2}}{2\sigma ^{2}}\right) 
\text{.}
\end{equation*}%
This distribution - a particular case of chi-square distribution $P(x)=\frac{%
\ x^{n-1}\exp (-x^{2}/2\sigma ^{2})}{2^{n/2-1}\sigma ^{n}\Gamma (n/2)}$,
with $n=2$ degrees of freedom - describes the sum of the squares of $n$
independent standard normal random variables. It is important to consider
that Rayleigh distribution occurs in many contexts of Physics, for instance,
in the spacing distribution of eigenvalues of symmetric random matrices and
in the known GOE (Gaussian orthogonal ensemble) to explain the spacing
distribution of energy levels in heavy nucleus \cite{MLMetha}. \ 

The idea of analysing the superposition of waves has motivated some authors
to hypothesize that constructive interference will lead to extreme events,
i.e., outliers events to a Rayleigh bulk. Rice in 1944, while exploring
aspects linked to the relationship between the energy spectrum of a surface
and its physical observables, considered that the elevation in a determined
point - in connection with analysis of the electrical noise current - is
given by a sum of different sine/cosine waves with different frequencies
with phases uniformly distributed. It was then obtained an
extension/generalization of Rayleigh distribution known as Rice distribution 
\cite{Rice}:%
\begin{equation*}
f(x\left\vert \nu ,\sigma \right. )=\frac{x}{\sigma ^{2}}\exp \left( -\frac{%
x^{2}+\nu ^{2}}{2\sigma ^{2}}\right) I_{0}\left (\frac{\nu x}{\sigma ^{2}}%
\right )
\end{equation*}%
where $I_{0}(x)$ is the 0-th order modified Bessel function of the first
kind. For large $x$, $I_{0}(x)$ can be approximated as $I_{0}(x)\approx 
\frac{e^{-x}}{\sqrt{2\pi x}}$, so that%
\begin{equation*}
f(x\left\vert \nu ,\sigma \right. )=\frac{(x/\nu )^{1/2}}{\sqrt{2\pi \sigma
^{2}}}\exp \left( -\frac{(x-\nu )^{2}}{2\sigma ^{2}}\right)
\end{equation*}%
which in 0-th order is given by a Gaussian distribution: $f(x\left\vert \nu
,\sigma \right. )\approx \frac{1}{\sqrt{2\pi \sigma ^{2}}}\exp \left( -\frac{%
(x-\nu )^{2}}{2\sigma ^{2}}\right) $.

More recently, several authors have been shown the existence of non-Rayleigh
effects in many different contexts. In a recent contribution the authors 
\cite{Bonatto2019} based on previous results \cite{Fisher2015} show that a
\textquotedblleft sudoku light phases sequence\textquotedblright\ which is
built from a set of memory rules to correlate the phases can generate light
rogue waves, i.e., intensities that large deviate from Rayleigh
distribution. We believe that a simpler and more general correlation
mechanism is enough to generate rogue waves, since the correlation
coefficient seems to be the main parameter to reach this extreme events.

The question here is then if we can observe rogue waves/extreme events by
controlling a correlation coefficient. As will be observed, the answer is
positive! So in this work, we propose a simple and didactic method to show
how the correlations imply deviation from Rayleigh distribution, by directly
controlling the phases correlation. We show here that extreme events occur
for both cases: discrete and continuous random variables.

In section \ref{Sec:Model} we present the details to generate correlated
random variables from independent random variables keeping the same variance
and we present a way to standardize our study for the different versions of
distributions: discrete and continuous ones. We considered uniform random
variables since such distribution has compact support, and we can control
with absolute accuracy, where the random phases will be generated in $[-\pi
,\pi ]$.

Our main results are presented in section \ref{Sec:Results} showing how the
rogue waves can emerge with the introduction of correlated phases. We
propose a simple way to measure the rogue-wave level obtained for different
correlations. Finally, we summarize our results in section \ref%
{Sec:Conclusions} where we also present our main conclusions.

\section{The model}

\label{Sec:Model}

Let us consider a superposition of waves corresponding to electric fields on
the Fraunhofer plane (far-field):%
\begin{equation}
\begin{array}{lll}
E & = & A_{\beta }\sum_{j=1}^{N}e^{i(\beta j+\phi _{j})} \\ 
&  &  \\ 
& = & A_{\beta }\sum_{j=1}^{N}\cos (\beta j+\phi _{j})+iA_{\beta
}\sum_{j=1}^{N}\sin (\beta j+\phi _{j})\text{,}%
\end{array}%
\end{equation}%
where $A_{\beta }=E_{0}\ $sinc$(\beta /2)$, where $E_{0}$ is the amplitude
of incident electric field, $\phi _{j}$ is a random phase of pixel $j$, and $%
\beta $ is related to the diffraction angle, and sinc$(x)=\frac{\sin x}{x}$.

We can calculate the intensity $I=\left\vert E\right\vert ^{2}$, thus 
\begin{equation*}
\begin{array}{lll}
I & = & A_{\beta }^{2}\left\{ \left[ \sum_{j=1}^{N}\cos (\beta j+\phi _{j})%
\right] ^{2}+\left[ \sum_{j=1}^{N}\sin (\beta j+\phi _{j})\right]
^{2}\right\} \\ 
&  &  \\ 
& = & A_{\beta }^{2}\left[ N+\sum_{j\neq l}\cos \left( \beta (j-l)+\left(
\phi _{j}-\phi _{l}\right) \right) \right]%
\end{array}%
\end{equation*}

Expanding the sum, we have:%
\begin{equation*}
\begin{array}{l}
\sum_{j\neq l}\cos \left( \beta (j-l)+\left( \phi _{j}-\phi _{l}\right)
\right) = \\ 
\\ 
=\sum_{j\neq l}\cos \beta (j-l)\left( \cos \phi _{j}\cos \phi _{l}+\sin \phi
_{j}\sin \phi _{l}\right) + \\ 
-\sum_{j\neq l}\sin \beta (j-l)\left( \sin \phi _{j}\cos \phi _{l}-\sin \phi
_{l}\cos \phi _{j}\right)%
\end{array}%
\end{equation*}

If the phases are independent random variables and identically distributed
according to a pdf $p\left( \phi \right) $, we can write (after some
algebra) and by the fact that $\sum_{j\neq l}\sin \beta (j-l)=0$ 
\begin{equation*}
\left\langle \sum_{j\neq l}\cos \left( \beta (j-l)+\left( \phi _{j}-\phi
_{l}\right) \right) \right\rangle =2(a^{2}+b^{2})\sum_{j<l}\cos \beta (j-l)
\end{equation*}%
where $a=\int_{\phi _{\min }}^{\phi _{\max }}d\phi p(\phi )\cos \phi $, and $%
b=\int_{\phi _{\min }}^{\phi _{\max }}d\phi p(\phi )\sin \phi $, where are
supposing that $\phi _{\min }$ and $\phi _{\max }$ are the extremal angles
of the distribution.

Performing the sum, we have

\begin{equation}
\left\langle \sum_{j\neq l}\cos \left( \beta (j-l)+\left( \phi _{j}-\phi
_{l}\right) \right) \right\rangle =(a^{2}+b^{2})\left[ \frac{\sin ^{2}(\frac{%
\beta N}{2})}{\sin ^{2}(\frac{\beta }{2})}-N\right]
\end{equation}

And therefore the intensities $I$ on the screen are given by:%
\begin{equation}
\left\langle I\right\rangle =A_{\beta }^{2}\left[ N+(a^{2}+b^{2})\left( 
\frac{\sin ^{2}(\frac{\beta N}{2})}{\sin ^{2}(\frac{\beta }{2})}-N\right) %
\right]  \label{Eq:Average}
\end{equation}

Thus, in this paper, we will study the probability density function (PDF), $%
P(I)$, according to the phases distribution $p(\phi )$ which are not
independent random variables, so that we can analyze the tail of $P(I)$ and
deviations in relation to $\left\langle I\right\rangle $. Moreover we also
analyze the possible effects on $P(I)$ considering that phases are discrete
random variables with continuous limit to a uniform distribution.

\subsection{Correlated random phases from non-correlated random phases}

In this section we will show that we can generate correlated random
variables from non-correlated random phases considering that both
(correlated and non-correlated) have the same variance and average. Let us
consider two random variables 
\begin{equation}
\begin{array}{ccc}
\phi _{1} & = & \alpha _{1}\varphi _{1}+\alpha _{2}\varphi _{2} \\ 
&  &  \\ 
\phi _{2} & = & \beta _{1}\varphi _{1}+\beta _{2}\varphi _{2}%
\end{array}%
\end{equation}%
where $\varphi _{1}$ and $\varphi _{2}$ are i.i.d. random variables, which
means: $\left\langle \varphi _{1}\right\rangle =\left\langle \varphi
_{2}\right\rangle =\left\langle \varphi \right\rangle $, and $\left\langle
\varphi _{1}\varphi _{2}\right\rangle =\left\langle \varphi
_{1}\right\rangle \left\langle \varphi _{2}\right\rangle =\left\langle
\varphi \right\rangle ^{2}$. The variance of variable $\phi _{1}$, for
example, can be calculated - after some cancellations - according to 
\begin{equation}
\begin{array}{l}
\left\langle \left( \Delta \phi _{1}\right) ^{2}\right\rangle =\left\langle
\phi _{1}^{2}\right\rangle -\left\langle \phi _{1}\right\rangle ^{2} \\ 
\\ 
=\left\langle (\alpha _{1}\varphi _{1}+\alpha _{2}\varphi
_{2})^{2}\right\rangle -\left\langle \alpha _{1}\varphi _{1}+\alpha
_{2}\varphi _{2}\right\rangle ^{2} \\ 
\\ 
=(\alpha _{1}^{2}+\alpha _{2}^{2})\left\langle \left( \Delta \varphi \right)
^{2}\right\rangle%
\end{array}%
\end{equation}%
where $\left\langle \varphi _{1}^{2}\right\rangle -\left\langle \varphi
_{1}\right\rangle ^{2}=\left\langle \varphi _{2}^{2}\right\rangle
-\left\langle \varphi _{2}\right\rangle ^{2}=\left\langle \left( \Delta
\varphi \right) ^{2}\right\rangle $, and similarly%
\begin{equation}
\left\langle \left( \Delta \phi _{2}\right) ^{2}\right\rangle =(\beta
_{1}^{2}+\beta _{2}^{2})\left\langle \left( \Delta \varphi \right)
^{2}\right\rangle \text{.}
\end{equation}%
Now we want the condition 
\begin{equation}
\left\langle \left( \Delta \phi _{1}\right) ^{2}\right\rangle =\left\langle
\left( \Delta \phi _{2}\right) ^{2}\right\rangle =\left\langle \left( \Delta
\varphi \right) ^{2}\right\rangle ,  \label{Eq:equality_of_dispersions}
\end{equation}%
which implies that $\alpha _{1}^{2}+\alpha _{2}^{2}=\beta _{1}^{2}+\beta
_{2}^{2}=1$.

It is worth noting that although $\varphi _{1}$ and $\varphi _{2}$ are not
non-correlated random variables, $\phi _{1}$ and $\phi _{2}$ are, and the
correlation between these random variables can be calculated: 
\begin{equation}
\rho =\frac{\left\langle \left( \ \phi _{1}-\left\langle \phi
_{1}\right\rangle \right) \left( \ \phi _{2}-\left\langle \phi
_{2}\right\rangle \right) \right\rangle }{\sqrt{\left\langle \left( \Delta
\phi _{1}\right) ^{2}\right\rangle \left\langle \left( \Delta \phi
_{2}\right) ^{2}\right\rangle }}
\end{equation}

Thus, again after some cancellations and combinations:

\begin{equation}
\begin{array}{ccc}
\left\langle \left( \ \phi _{1}-\left\langle \phi _{1}\right\rangle \right)
\left( \ \phi _{2}-\left\langle \phi _{2}\right\rangle \right) \right\rangle
& = & \left\langle \ \phi _{1}\phi _{2}\right\rangle -\ \left\langle \phi
_{1}\right\rangle \left\langle \phi _{2}\right\rangle \\ 
&  &  \\ 
& = & (\alpha _{1}\beta _{1}+\alpha _{2}\beta _{2})\left\langle \left(
\Delta \varphi \right) ^{2}\right\rangle%
\end{array}%
\end{equation}
$\ $

Thus, we can conclude:

\begin{equation}
\rho =(\alpha _{1}\beta _{1}+\alpha _{2}\beta _{2})
\end{equation}

Given the properties previously considered, we can denote $\alpha _{1}=\beta
_{2}=\sin \theta $ and $\alpha _{2}=\beta _{1}=\cos \theta $, and therefore $%
\sin 2\theta =\rho $, so that 
\begin{equation}
\theta =\frac{1}{2}\sin ^{-1}(\rho )
\end{equation}

Thus the random variables%
\begin{equation}
\phi _{1}=\sin \left( \frac{1}{2}\sin ^{-1}(\rho )\right) \varphi _{1}+\cos
\left( \frac{1}{2}\sin ^{-1}(\rho )\right) \varphi _{2}  \label{Eq:var1}
\end{equation}%
and 
\begin{equation}
\phi _{2}=\cos \left( \frac{1}{2}\sin ^{-1}(\rho )\right) \varphi _{1}+\sin
\left( \frac{1}{2}\sin ^{-1}(\rho )\right) \varphi _{2}  \label{Eq:var2}
\end{equation}%
have the same average that are given by:

\begin{equation*}
\begin{array}{ccccc}
\left\langle \phi _{1}\right\rangle & = & \left\langle \phi _{2}\right\rangle
& = & \frac{1}{\sqrt{2}}\left[ \left( 1-\sqrt{1-\rho ^{2}}\right)
^{1/2}+\left( 1+\sqrt{1-\rho ^{2}}\right) ^{1/2}\right] \left\langle \varphi
\right\rangle%
\end{array}%
\end{equation*}%
and the same dispersion of $\varphi _{1}$and $\varphi _{2}$ according to Eq.
(\ref{Eq:equality_of_dispersions}). It is important to emphasize that $\phi
_{1}$ and $\phi _{2}$ have the same variance of $\varphi _{1}$ and $\varphi
_{2}$ since $\left\langle \left( \Delta \phi _{1}\right) ^{2}\right\rangle
=\left\langle \left( \Delta \phi _{2}\right) ^{2}\right\rangle =\left\langle
\left( \Delta \varphi _{1}\right) ^{2}\right\rangle =\left\langle \left(
\Delta \varphi _{2}\right) ^{2}\right\rangle =\left\langle \left( \Delta
\varphi \right) ^{2}\right\rangle $, but the averages of $\phi _{1}$ and $%
\phi _{2}$ are not the same of $\varphi _{1}$ and $\varphi _{2}$ that are
equal to $\left\langle \varphi \right\rangle $, unless $\left\langle \varphi
\right\rangle =0$. So if one considers $\phi _{1}$ and $\phi _{2}$ as $\rho $%
-correlated random variables generated from two independent random variables 
$\varphi _{1}$ and $\varphi _{2}$ -- with average zero and variance $\sigma
^{2}=\left\langle \left( \Delta \varphi \right) ^{2}\right\rangle $--, $\phi
_{1}$ and $\phi _{2}$ also have average zero and variance $\sigma ^{2}$.

Here one has to be careful if we want to study the effects on the wave
amplitudes by introducing phase correlations, since if one works with
different average and dispersion between the cases $\rho =0$ and $\rho \neq
0 $\ the results can be misleading since there will be no parameter for a
fair comparison.

\subsection{Discrete and continuous random variables}

In this paper we consider phases which are uniformly distributed, since
there is no reason for some phases to be more probable than others. So to
study possible effects or similarities between discrete and continuous
version of this distribution on the $P(I)$ shape, we generalize our method
considering an equiprobable distribution for the phases, starting from a few
number of states $Q$ up to $Q\rightarrow \infty $ which corresponds to a
continuous and uniformly distributed random variable.

Our phases $\varphi _{j}$, $j=1,2,....,N$, must be taken from the interval $%
[-\pi ,\pi ]$ with symmetric PDF $p_{\rho }(\varphi )$, since $\left\langle
\varphi \right\rangle =0$. So we propose a simple discrete random variable:%
\begin{equation}
\varphi _{k}=\left( \frac{2k}{Q-1}-1\right) \pi  \label{Eq:fi}
\end{equation}%
where $Q>1$, an odd number of states by construction with $k=0,1,2...,Q-1$.
As an example, for $Q=3$ we have $\varphi _{0}=-\pi $, $\varphi _{1}=0$, $%
\varphi _{2}=\pi $. For $Q=5$, we have $\varphi _{0}=-\pi $, $\varphi
_{1}=-\pi /2$, $\varphi _{2}=0$, $\varphi _{3}=\pi /2$, $\varphi _{4}=\pi $.
If we define $p_{k}=p(\varphi =\varphi _{k})=\frac{1}{Q}$, i.e., all phases
occur with same probability, so we have 
\begin{equation}
\left\langle \varphi _{k}\right\rangle =\frac{\pi }{Q}\sum_{k=0}^{Q-1}\left( 
\frac{2k}{Q-1}-1\right) =0  \label{Eq:averfi}
\end{equation}%
and we want to calculate 
\begin{equation}
\begin{array}{lll}
\left\langle \varphi _{k}^{2}\right\rangle & = & \sum_{k=0}^{Q-1}\varphi
_{k}^{2}p_{k} \\ 
&  &  \\ 
& = & \frac{\pi ^{2}}{Q}\sum_{k=0}^{Q-1}\left( \frac{2k}{Q-1}-1\right) ^{2}
\\ 
&  &  \\ 
& = & \frac{\pi ^{2}}{Q}\left[ \frac{4}{(Q-1)^{2}}\sum_{k=0}^{Q-1}k^{2}%
\right. \\ 
&  & \left. -\frac{4}{(Q-1)}\sum_{k=0}^{Q-1}k+\sum_{k=0}^{Q-1}1\right] \\ 
&  &  \\ 
& = & \frac{\pi ^{2}}{3}\frac{Q+1}{Q-1}%
\end{array}
\label{Eq:sec_moment}
\end{equation}

The importance of this distribution is twofold:

\begin{enumerate}
\item The distribution is discrete and $p_{k}$ becomes the uniform
distribution when $Q\rightarrow \infty $. 
\begin{equation}
p(\varphi )=\left\{ 
\begin{array}{ccc}
\frac{1}{2\pi } & \text{if} & \left\vert \varphi \right\vert \leq \pi \\ 
0 & \text{if} & \left\vert \varphi \right\vert >\pi%
\end{array}%
\right.  \label{Eq:Uniform}
\end{equation}%
Also, $\left\langle \varphi _{k}\right\rangle =\left\langle \varphi
\right\rangle =0$, and from Eq. (\ref{Eq:sec_moment}), $\left\langle \varphi
_{k}^{2}\right\rangle =\left\langle \varphi ^{2}\right\rangle =\frac{\pi ^{2}%
}{3}$, i.e., we can study a system with discrete phases and analyze what
occur when $Q\rightarrow \infty $;

\item It is more interesting to study a distribution with compact support
for the phases since $-\pi \ \leq \ \varphi \ \leq \pi $. Thus, one makes
that the boundary is respected which results necessarily in considering $%
\sigma =\pi /\sqrt{3}.$
\end{enumerate}

In what follows, we present the results for the distribution of amplitudes $%
I $ considering random correlated phases given by Eqs. \ref{Eq:var1} and \ref%
{Eq:var2} from random independent variables drawn according to Eq. \ref%
{Eq:fi}, and also by studying the limit case ($Q\rightarrow \infty $)
according to Eq. \ref{Eq:Uniform}.

\section{Results}

\label{Sec:Results}

We computed the intensities distribution $I$, i.e., the diffraction
patterns. We used $N=1024$ random waves considering 300 values of $\beta $
equally spaced in the interval $[0,2\pi ]$. We repeated the procedure for an
ensemble of $N_{run}=10000$ realizations.

\begin{figure}[h]
\begin{center}
\includegraphics[width=1.0%
\columnwidth]{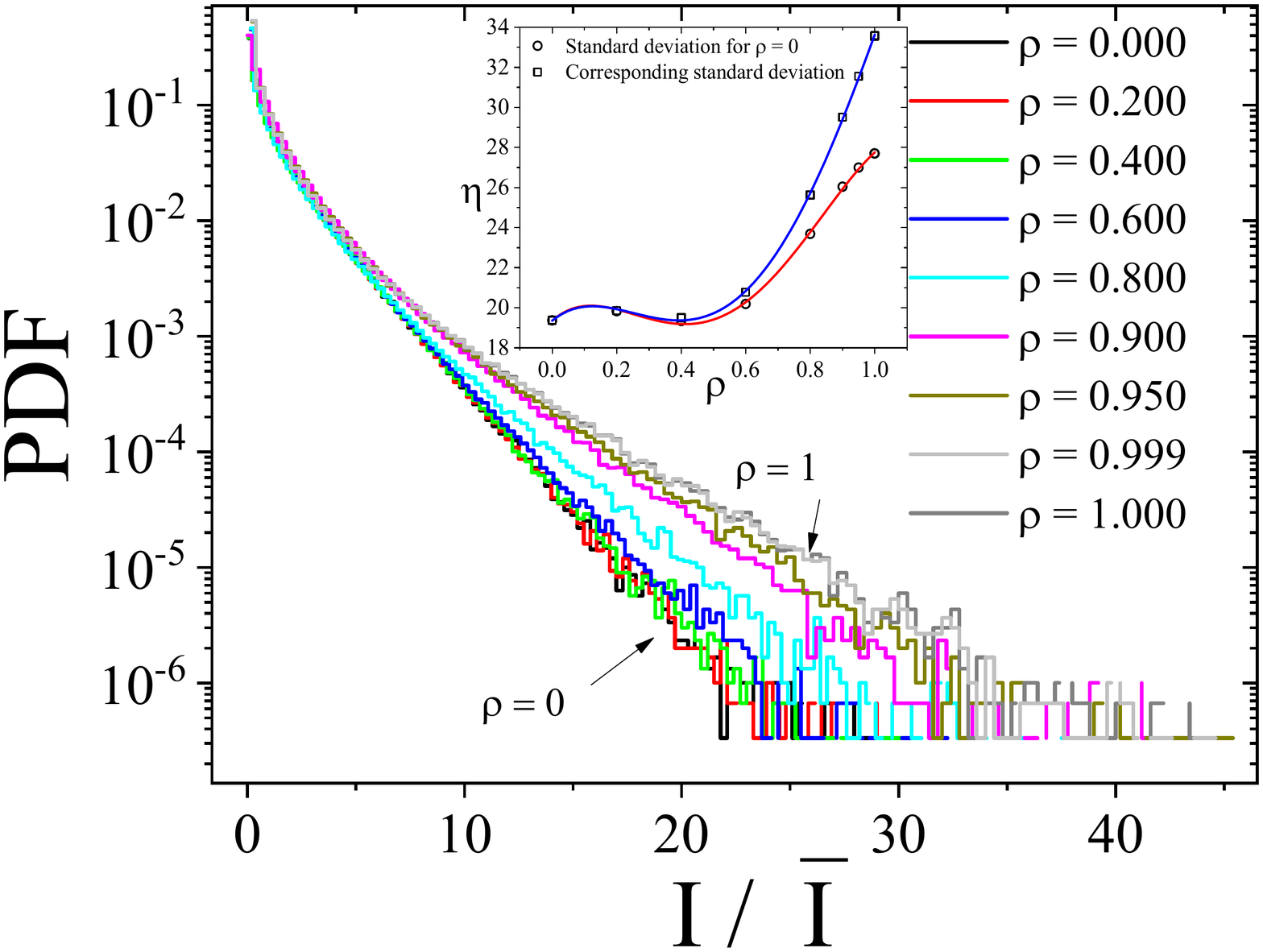}
\end{center}
\caption{Distribution of intensities for different values of $\protect\rho $
considering phases uniformly generated in $[-\protect\pi ,\protect\pi ]$
(shuffled case). The inset plot show the variable $\protect\eta =(I_{\max }-%
\overline{I})/\protect\sigma $ for each correlation. We can observe that $%
\protect\eta $ reaches 28 for $\protect\rho =1$. }
\label{Fig:PDF_shuffled_rogue_wave_factor}
\end{figure}

The $N=1024$ random phases generated for each realization are shuffled since
the correlated random variables are generated by pairs. Figure \ref%
{Fig:PDF_shuffled_rogue_wave_factor} shows the probability density function
(PDF) of light intensities scaled with total average light intensity sampled
(i.e., including all values of $\beta $). We used a mono-log scale. We can
observe that tails become heavier as $\rho $ increases indicating the
existence of rogue waves.

\begin{figure}[h]
\begin{center}
\includegraphics[width=0.5\columnwidth]{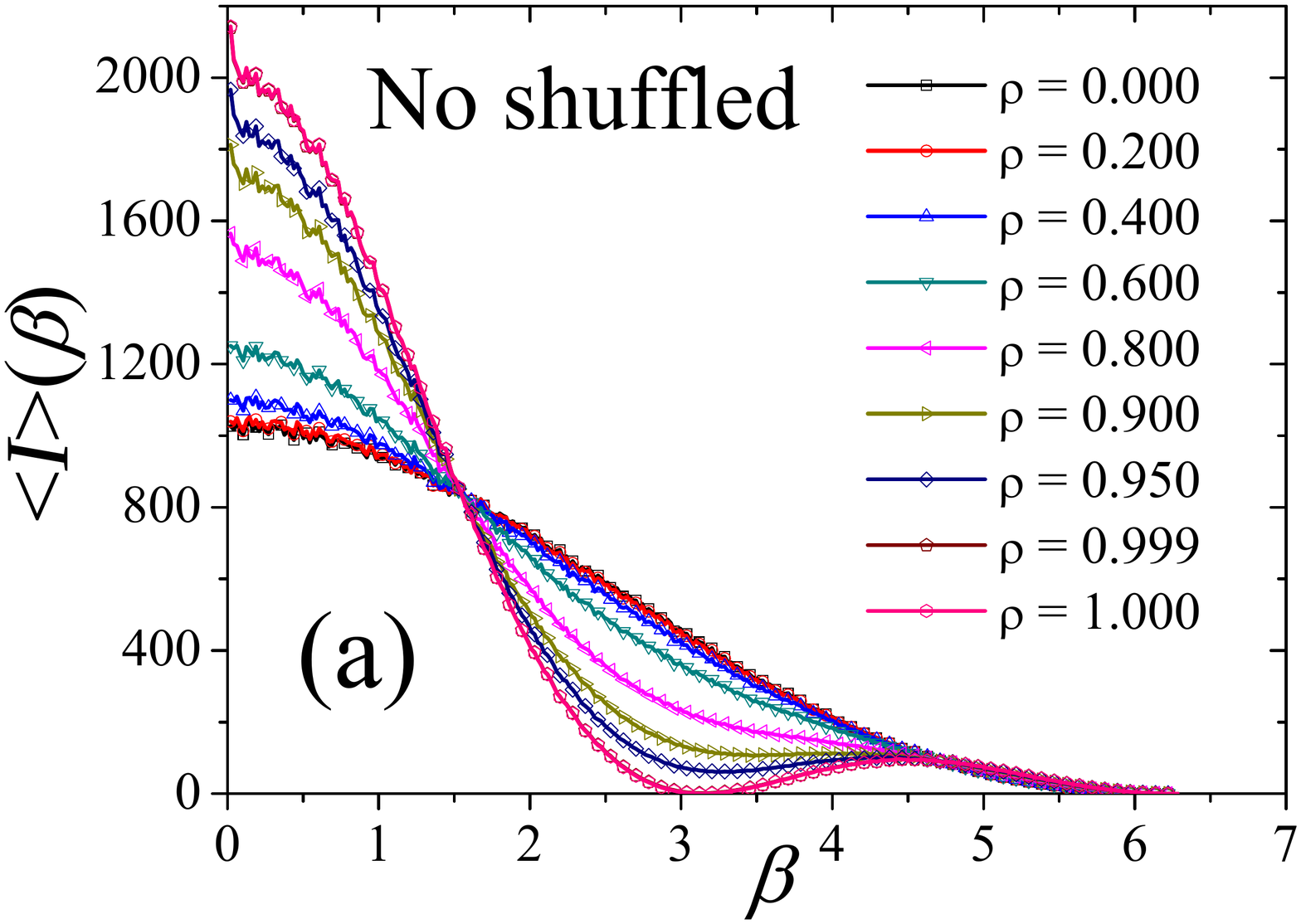}%
\includegraphics[width=0.5\columnwidth]{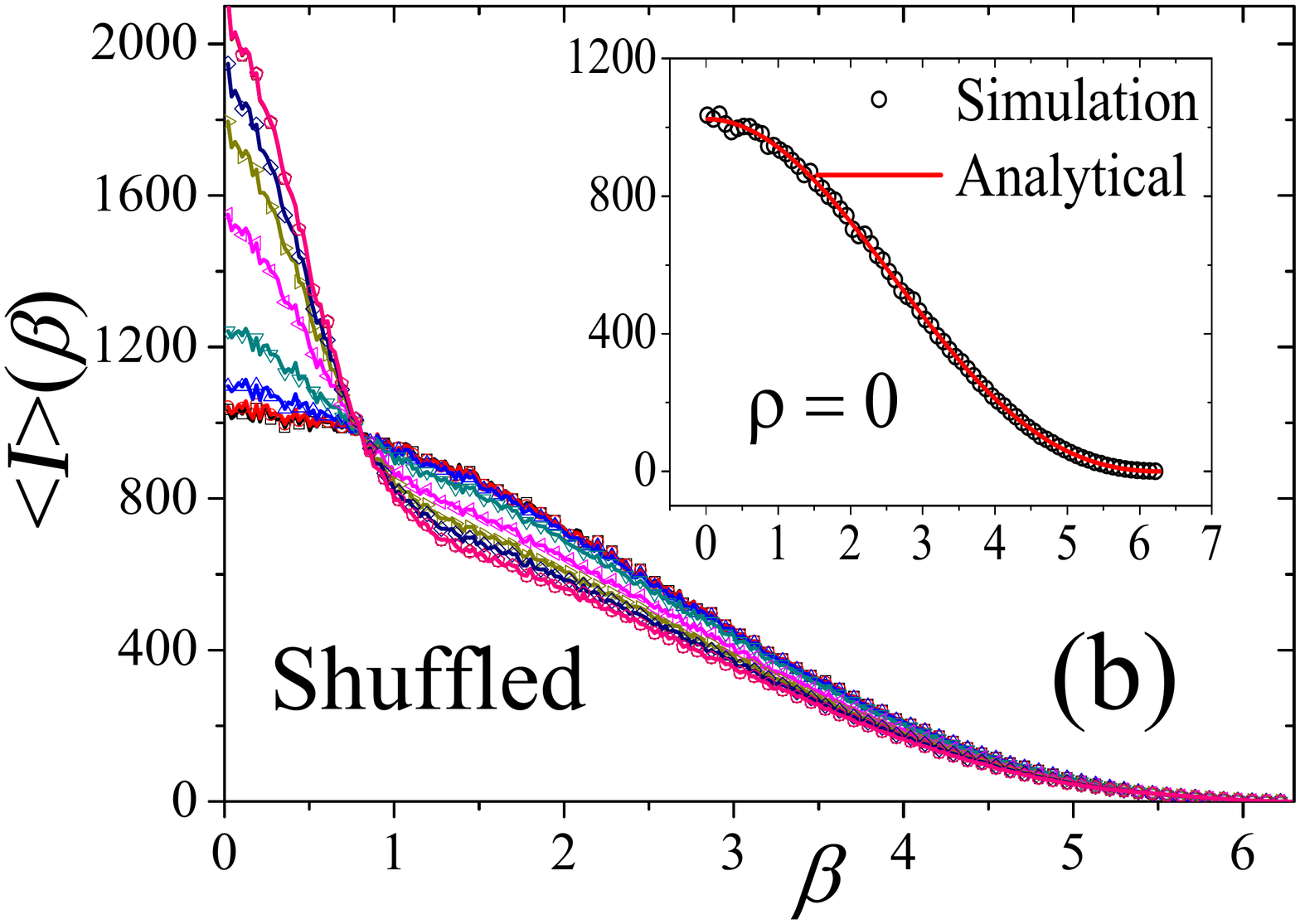} %
\includegraphics[width=0.5\columnwidth]{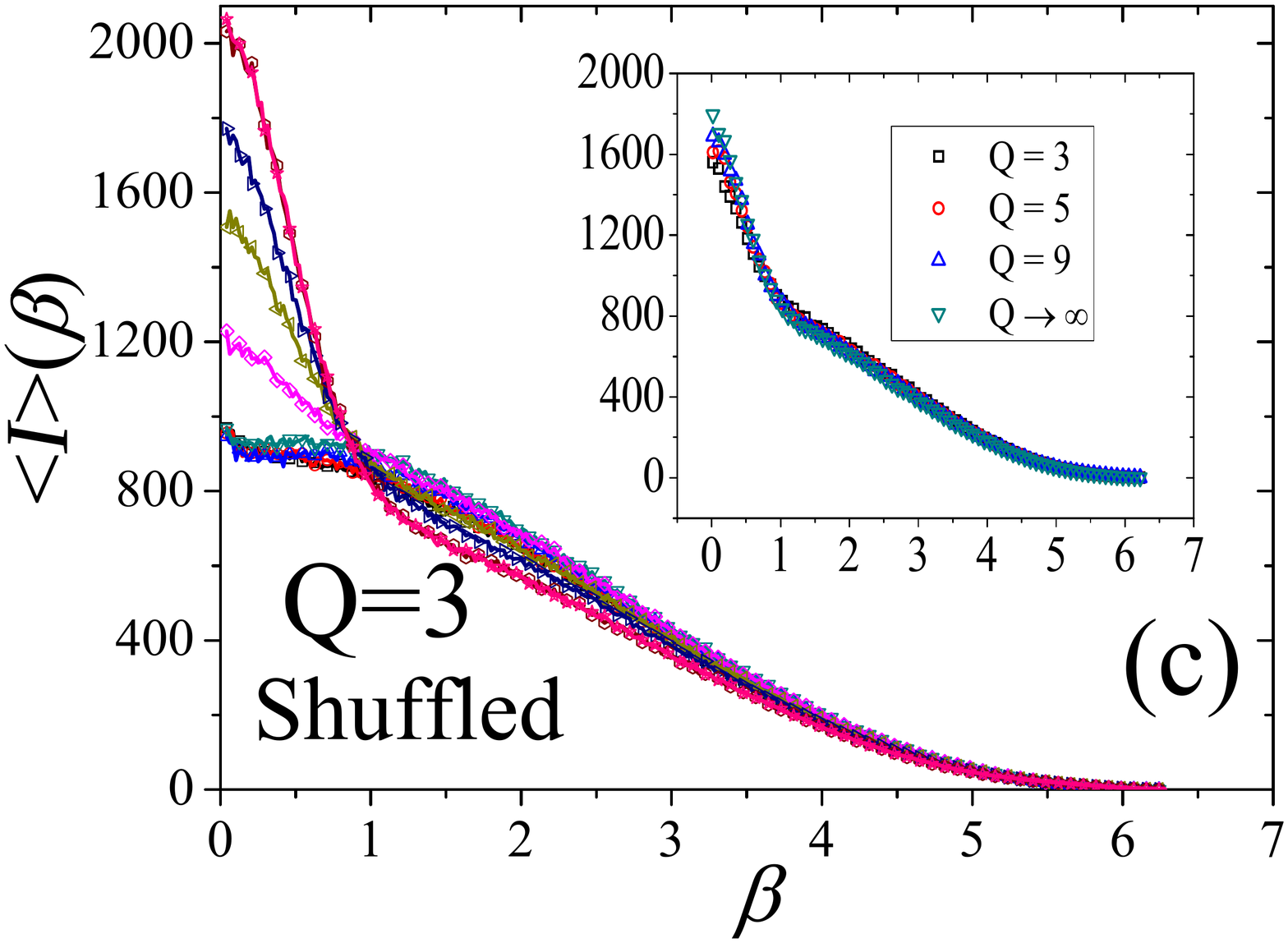}
\end{center}
\caption{(a) $\left\langle I\right\rangle $ as function of $\protect\beta $
for the shuffled uniform correlated phases distribution. (b) The same plot
for no shuffled random phases. Only as a double check, the inset plot shows
the agreement \ between the analytical $\left\langle I\right\rangle =NA_{%
\protect\beta }^{2}$ and results from simulations for $\protect\rho =0$. (c)
Effects of the discretization on the $\left\langle I\right\rangle $. }
\label{Fig:Average_intensity}
\end{figure}

For each value of $\rho $ one measures%
\begin{equation*}
\eta (\rho )=\frac{I_{\max }(\rho )-\overline{I}(\rho )}{\sigma (\rho )}
\end{equation*}%
which measures the rogue-wave level. The inset plot in Fig. \ref%
{Fig:PDF_shuffled_rogue_wave_factor} shows that for $\rho =1$, we have $\eta
=28$ while $\eta \approx 18$ for $\rho =0$ (open balls). However we should
calculate $\eta =\frac{I_{\max }(\rho )-\overline{I}(\rho )}{\sigma (\rho =0)%
}$, take into account the standard deviation in relation to our base: $\rho
=0$ (non correlated). In this case we can obtain $\eta \approx 34$.

We also calculated $\left\langle I\left( \beta \right) \right\rangle $ as
function of $\beta $ for different values of $\rho $. We performed two
different studies. First, we analyze the effects of not shuffling or
shuffling the sample of correlated random variables according to Figs. \ref%
{Fig:Average_intensity} (a) and (b). In Fig. \ref{Fig:Average_intensity}
(b), the inset plot shows that results for $\rho =0$, according to Eq. \ref%
{Eq:Average} (where $a=b=0$ according with our distribution) perfectly
agrees with simulation results.

After, we study the effects of the discretization of random variables on $%
\left\langle I\left( \beta \right) \right\rangle $ as observed in Fig. \ref%
{Fig:Average_intensity} (c). We do not have sensitive differences in
comparison with results obtained with continuous distribution described in
Fig. \ref{Fig:Average_intensity} (b). The inset plot in Fig. \ref%
{Fig:Average_intensity} (c) shows that regardless of $Q\ $($Q\geq 3$) the
behavior is similar to that one obtained in \ref{Fig:Average_intensity} (b),
which shows that discretization is not indeed an important factor in the
average $\left\langle I\left( \beta \right) \right\rangle $.

\begin{figure}[h]
\begin{center}
\includegraphics[width=1.0\columnwidth]{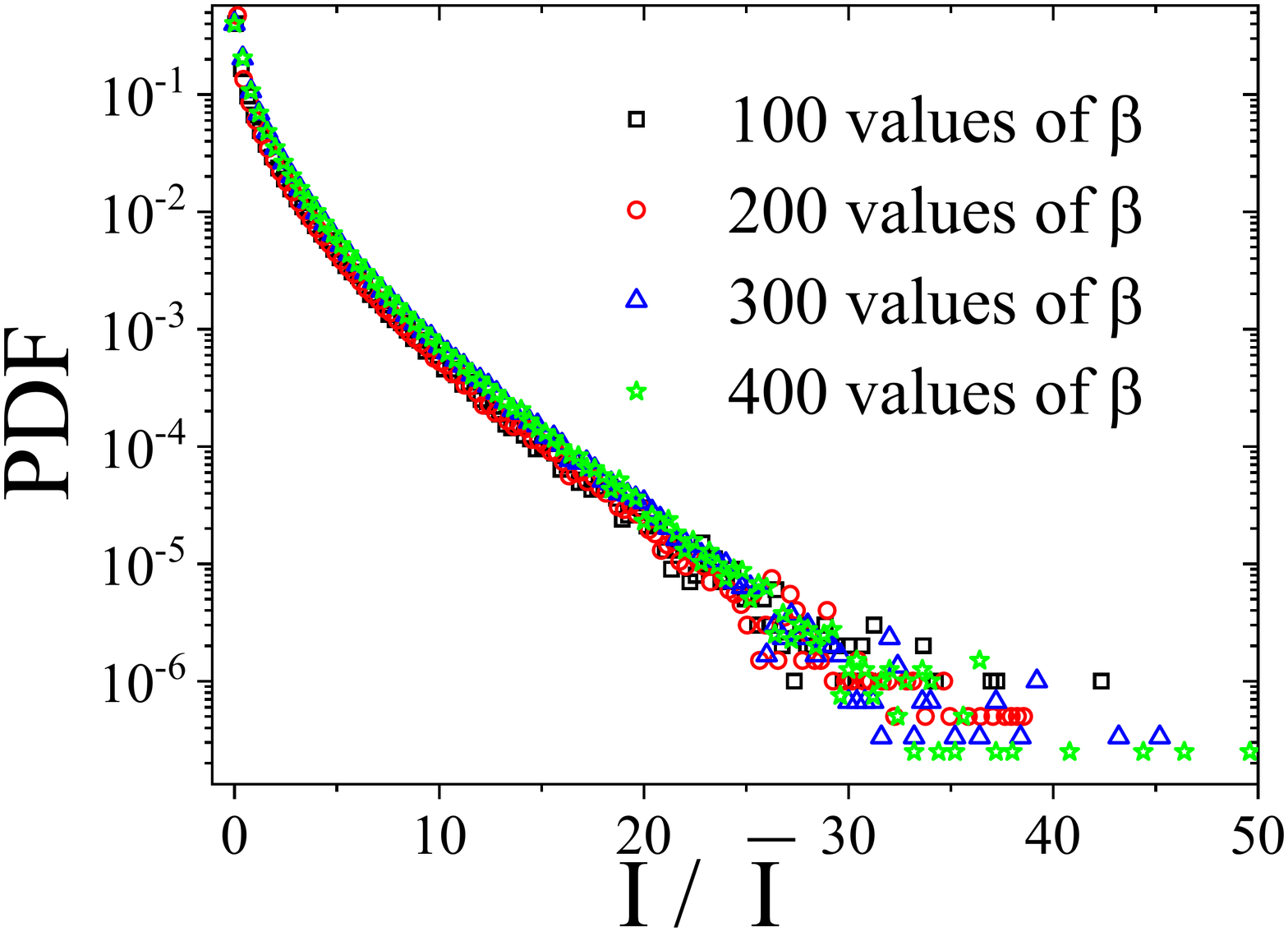}
\end{center}
\caption{PDF considering different number of repartitions of the interval $%
[0,2\protect\pi ]$ for $\protect\beta $. }
\label{Fig.Effects_of_Nbeta}
\end{figure}

Finally, we analyzed the effects of the partitioning of the interval $%
[0,2\pi ]$ for $\beta $, since the PDFs of the intensities should depend on
the granulation of the inverval. So we partitioned the interval in $N_{\beta
}=100,200,300,$and $400$ parts, and we calculated the PDF for all these
cases. These results are shown in Fig. \ref{Fig.Effects_of_Nbeta}.

We used 300 subdivisions for the interval $[0,2\pi ]$ in all previous cases
studied in this manuscript. We can observe that there is no visual
difference between $N_{\beta }=300$ and $400$, and qualitatively we have no
differences for all studied cases, by showing that we need no greater values
of $N_{\beta }\ $in order to build the histograms of PDFs in our results.

\section{Summaries, conclusions, and discussions}

\label{Sec:Conclusions}

In this work we show the existence of rogue waves for PDF of the intensities
associated to the superposition of waves of electric fields with the
introduction of random correlated phases controlled by the correlation
coefficient $\rho $. We can observe that heavier tail distributions are
obtained as the correlation increases. Our results also suggest that
discretization is not an important ingredient to change the PDF shape. The
average intensity is also studied as function of $0\leq \beta \leq 2\pi $,
which corroborates the analytical results. Shuffling procedure leads to a
reduction of the oscillatory effects on the $\left\langle I\right\rangle
(\beta )$ (see Eq. \ref{Eq:Average}). Our results are simple and easy to be
applied, and deserve future investigations in other problems involving
extreme statistics.

\textbf{Acknowledgements}

R. da Silva were financially supported by CNPq under grant numbers:
311236/2018-9, 424052/2018-0. The authors thank R. Rego, C. Bonatto IF-UFRGS
for the fruitful discussions.


\begin{thebibliography}{9}
\bibitem{Onorato} M. Onorato, T. Waseda, A. Toffoli, L. Cavaleri, O.
Gramstad, P.A.E.M. Janssen, T. Kinoshita, J. Monbaliu, N. Mori, A.R.
Osborne, M. Serio, Phys. Rev. Lett., \textbf{102} 114502 (2009).

\bibitem{Dudley2014} J. M. Dudley, F. Dias, M. Erkintalo, G. Gentry, Nature
Photonics, \textbf{8} 755--764 (2014)

\bibitem{YZhenYa2010} Y. Zhen-Ya, Comm. Theor. Phys. \textbf{54\ }947--949
(2010)

\bibitem{Koussaifi} R. El Koussaifi, A. Tikan, , A. Toffoli, S. Randoux, P.
Suret, and M. Onorato, Phys. Rev. E, \textbf{97} 012208 (2018)

\bibitem{MLMetha} M. L. Metha, Random Matrices and the Statistical Theory of
Energy Levels, Academic Press (1967)

\bibitem{Rice} S.O. Rice , Bell Syst. Tech. \textbf{23}, 282--332 (1944).

\bibitem{Bonatto2019} C. Bonatto, S. D. Prado, F. L. Metz, Julio, R.
Schoffen,1 R. R. B. Correia, J. M. Hickmann (submitted to Phys. Rev. Lett.)

\bibitem{Fisher2015} R. Fischer, I. Vidal, D. Gilboa, R. R. B. Correia, A.
T. Ribeiro-Teixeira, S. D. Prado, J. Hickman, Y. Silberberg, Phys. Rev.
Lett. \textbf{115}, \ 073901 (2015).
\end{thebibliography}
\end{document}